\documentclass[a4paper,one]{amsart}
\usepackage{amsthm}
\usepackage{amsmath}
\usepackage{amsfonts}
\usepackage{amssymb}
\usepackage{mathrsfs}
\usepackage{hyperref}
\usepackage[numbers]{natbib}
\usepackage{fancyhdr}
\usepackage{color}
\theoremstyle{plain}
\newtheorem{lem}{Lemma}
\newtheorem{prop}[lem]{Proposition}
\newtheorem{thm}[lem]{Theorem}
\newtheorem{cor}[lem]{Corollary}

\theoremstyle{definition}

\newtheorem{rem}[lem]{Remark}
\newcommand{\R}{\mathbb R}
\newcommand{\Z}{\mathbb Z}
\newcommand{\N}{\mathbb N}
\newcommand{\Diff}{\mbox{\rm Diff}}

\newcommand{\id}{\text{\rm id}}
\newcommand{\dx}{\,\text{\rm d}x}
\renewcommand{\d}{\,\text{\rm d}}

\newcommand{\g}{\mathfrak{g}}
\newcommand{\Ad}{\text{\rm Ad}}
\newcommand{\ad}{\text{\rm ad}}
\renewcommand{\S}{\mathbb S}
\renewcommand{\phi}{\varphi}

\newcommand{\eps}{\varepsilon}
\newcommand{\ska}[2]{\left\langle #1,#2\right\rangle}
\newcommand{\set}[2]{\left\{#1;\;#2\right\}}
\newcommand{\bea}{\begin{eqnarray}}
\newcommand{\eea}{\end{eqnarray}}
\newcommand{\beq}{\begin{equation}}
\newcommand{\eeq}{\end{equation}}
\renewcommand{\autoref}[1]{\text{Eq.}~\eqref{#1}}
\newcommand{\red}[1]{\color{black}#1\color{black}}
\begin{document}
\title{The curvature of semidirect product groups associated with \red{two-component Hunter-Saxton systems}}
\author{Martin Kohlmann}
\address{Institute for Applied Mathematics, University of Hannover, D-30167 Hannover, Germany}
\email{kohlmann@ifam.uni-hannover.de}
\date{\today}
\begin{abstract}
In this paper, we study two-component versions of the periodic
Hunter-Saxton equation and its $\mu$-variant. Considering both
equations as a geodesic flow on the semidirect product of the
circle diffeomorphism group $\Diff(\S)$ with a space of scalar
functions on $\S$ we show that both equations are locally
well-posed. The main result of the paper is that the sectional
curvature associated with the 2HS is constant and positive and
that 2$\mu$HS allows for a large subspace of positive sectional
curvature. The issues of this paper are related to some of the
results for 2CH and 2DP presented in \cite{EKL10}.
\end{abstract}
%
%
%
%
\maketitle
\section{Introduction}
The Hunter-Saxton (HS) equation
\bea\label{HS}u_{txx}+2u_xu_{xx}+uu_{xxx}=0,\quad x\in\S,\quad
t>0,\eea
was derived in \cite{HS91} as a model for propagation of
orientation waves in a massive nematic liquid crystal director
field. The function $u(t,x)$ depends on a space variable $x$ and a
slow time variable $t$.
\red{In recent years, the question of the existence and regularity of solutions to \eqref{HS} as well as integrability properties have been examined in great detail, cf., e.g., \cite{BSS01,C05,HZ94,Y04}. }
The HS equation can be regarded as the
$\alpha\to\infty$-limit of the Camassa-Holm (CH) equation
$$m_t=-(m_xu+2mu_x),\quad m=(1-\alpha\partial_x^2)u=u-\alpha u_{xx}$$
which was introduced to model the shallow-water medium-amplitude
regime for wave motion over a flat bad, \cite{CH93}; more
precisely, the HS equation is equivalent to
\bea m_t=-(m_xu+2mu_x),\quad
m=(-\partial_x^2)u=-u_{xx}.\label{HS'}\eea
\red{Alternatively, the HS equation can be regarded as the high-frequency
or short-wave limit $(x,t)\mapsto(\eps x,\eps t)$, for $\eps\to 0$,
of the CH equation, cf.~\cite{DP98,HZ94}. } Similarly to the CH, the HS equation comes up from Langrange's
variational principle with the Lagrangian
$$\mathcal L_{HS}(u)=\frac{1}{2}\int_\S u_x^2\dx,$$
\red{cf.~\cite{HZ94}, which differs from the Lagrangian for CH only in a term proportional to $u^2$ under the integral sign}. The Euler-Lagrange equation for the modified Lagrangian
$$\mathcal L_{2HS}(u,\rho)=\frac{1}{2}\int_\S u_x^2\dx+\frac{1}{2}\int_\S\rho^2\dx$$
is the 2HS equation
\bea\label{2HS}\left\{\begin{array}{lcl}
  m_t & = & -m_xu-2mu_x-\rho\rho_x,\\
    \rho_t & = & -(\rho u)_x, \\
\end{array}\right.\eea
an integrable two-component extension of \eqref{HS} which reduces
to \eqref{HS} for $\rho=0$.

\red{ We note that the two-component HS equation is a particular case of the Gurevich-Zybin system pertaining to nonlinear one-dimensional dynamics of dark matter as well as nonlinear ion-acoustic waves, cf.~\cite{P05}. Also, it is related to the two-component Camassa-Holm (2CH) system (which reads \eqref{2HS} with $m=u-u_{xx}$, \cite{CI08,ELY07'}) via the short-wave limit. Note that for the 2CH equation, the second variable $\rho$ can be interpreted as the fluid's density and its evolution equation in fact is the continuity equation for velocity and fluid density. The two-component Hunter-Saxton \eqref{2HS} equation is a special case of the two-parameter family of evolution equations
\bea
\left\{
\begin{array}{rcl}
m_t  & = & -m_xu-au_xm-\kappa\rho\rho_x,\\
m    & = & -u_{xx},\\
\rho & = & -\rho u_x+(1-a)u_x\rho,
\end{array}
\right.
\quad
(a,\kappa)\in\R^2,
\label{general2HS}\eea
from which it emerges for the choice $(a,\kappa)=(2,1)$.
The system \eqref{general2HS} is of great importance for a variety of problems occurring in mathematical physics: A special case of the two-parameter family \eqref{general2HS} is the Proudman-Johnson equation \cite{O09,PJ62} for $\rho=0$ and $a=-1$; this equation is obtained from the incompressible 2D Euler equations by a special ansatz for the stream function. Further relations to, e.g., the inviscid K\'arm\'an-Batchelor flow \cite{Ch08,HL08}, which derives from the 3D axisymmetric Euler equations, or the famous Constantin-Lax-Majda equation \cite{CLM85} are explained in \cite{W10}. We also refer the reader to \cite{W10} for an extensive discussion of well-posedness and blow-up for the general system \eqref{general2HS}.

What makes the 2HS system particularly interesting is its potential exhibition of nonlinear phenomena as breaking waves and peakons \cite{CI08}.
}
In \cite{LL09} the authors derive the
2HS equation as the $N=2$ supersymmetric extension of the
Camassa-Holm equation. They also work out the bi-Hamiltonian
formulation and a Lax pair representation for the 2HS equation.\
Concerning geometry, the 2HS equation can be regarded as an Euler
equation on the superconformal algebra of contact vector fields on
the $1|2$-dimensional supercircle. Finally, \cite{LL09} show some
explicit solutions of \eqref{2HS}, like bounded traveling waves.

Interesting variants of Eqs.~\eqref{HS} and \eqref{2HS} are
obtained by \red{using $m=\mu(u)-u_{xx}$ instead of $m=-u_{xx}$ in \eqref{HS'} and \eqref{2HS}, }where $\mu(u)$ is the mean
of $u:\S\to\R$. This way we obtain the $\mu$HS equation
\beq 0=u_{txx}+2u_xu_{xx}+uu_{xxx}-2\mu(u)u_x\label{muHS}\eeq
and its two-component extension
\bea\label{2muHS}\left\{\begin{array}{lcl}
  0&=&u_{txx}+2u_xu_{xx}+uu_{xxx}-2\mu(u)u_x-\rho\rho_x,\\
  \rho_t & = & -(\rho u)_x. \\
\end{array}\right.\eea
\red{In \cite{KLM08} it is explained that the $\mu$HS equation models nematic liquid crystals with a preferred direction of the director field, e.g., coming from an external magnetic field acting on dipoles and turning them to this direction. Our motivation for studying the 2-component extension $2\mu$HS comes from the observation that the generalized system possesses an integrable structure \cite{Z10} and the expectancy that similar relationships between $\mu$HS and $2\mu$HS compared to HS and 2HS might hold true. }
Both, \eqref{muHS} and \eqref{2muHS}, are integrable equations
related to the isospectral problem
\bea\left\{
\begin{array}{ccc}
  \psi_{xx}+(m\lambda+\rho^2\lambda^2)\psi & = & 0,\\
  \psi_t & = & -\left(\frac{1}{2\lambda}+u\right)\psi_x+\frac{1}{2}\psi
u_x, \\
\end{array}%
\right.\nonumber\eea
i.e., $\psi_{txx}=\psi_{xxt}$ and $\lambda_t=0$ imply the 2$\mu$HS
equation \eqref{2muHS}.
Equations \eqref{muHS} and \eqref{2muHS} follow from the
variational principle with the Lagrangians
$$\mathcal L_{\mu HS}(u)=\frac{1}{2}\int_\S u(\mu-\partial_x^2)u\dx$$
and
$$\mathcal L_{2\mu HS}(u,\rho)=\frac{1}{2}\int_\S u(\mu-\partial_x^2)u\dx+\frac{1}{2}\int_\S\rho^2\dx.$$
\red{While there are some well-posedness results for the $\mu$HS in \cite{KLM08}, there is no fully developed existence and regularity theory for its two-component version up to now. A first attempt to construct global weak solutions is presented in \cite{LY11}.

We now turn to the geometric theory behind \eqref{HS}, \eqref{2HS}, \eqref{muHS} and \eqref{2muHS} which will be relevant for this paper: }
Equations \eqref{HS} and \eqref{muHS} have been studied as
geodesic flow on the circle diffeomorphism group $H^s\Diff(\S)$ of Sobolev class $H^s$, \cite{L07',KLM08}.
\red{We provide some elementary results about recasting Euler equations as geodesic equations on infinite-dimensional Lie groups in an appendix of this paper; see Section~\ref{sec_appendix}. }
The
strategy is to define a right-invariant metric on $H^s\Diff(\S)$,
induced at the identity by the inertia operators $-\partial_x^2$
and $\mu-\partial_x^2$ respectively, and to prove its
compatibility with an affine connection which is given canonically
in terms of the Christoffel operator for the respective
equation. This implies the existence of a geodesic flow and first
local well-posedness results\footnote{For HS one has to factorize
$H^s\Diff(\S)$ by a subgroup as we will explain in the
following.}. Moreover, computations of the curvature tensor and
the sectional curvature for both equations have been performed.
The meaningful results are that the sectional curvature is
strictly positive for both equations and that HS has constant
positive sectional curvature which motivates that, in this case,
$H^s\Diff(\S)$ is locally isometric to an $L_2$-sphere. This
sphere interpretation has various geometric consequences as
explained in \cite{L07'',L08}. Furthermore, the positivity of sectional
curvatures is related to stability issues for the
geodesics\red{, cf. Section~\ref{sec_appendix}}.

In this paper, we extend some of the above results to the
integrable 2-component extensions \eqref{2HS} and \eqref{2muHS}.
In a first step we comment on how to choose a suitable
configuration manifold for 2HS and 2$\mu$HS; here, we will make
use of semidirect products. Then we give a proof of the fact that
the curvature tensor for 2HS is of the same form as for HS with
the only difference that the geometric objects contained therein
have to be replaced by their two-component analogs. For 2$\mu$HS
we first obtain a curvature formula which is similar to the
formula for the one-component $\mu$HS. Second, we compute the
sectional curvature in several two-dimensional directions and
establish a positivity result. The paper is organized as follows:

Section 2 explains the geometric structure of 2HS and 2$\mu$HS;
here we first recall some results about the corresponding one-component
equations and semidirect products which we will need in the following. The main
goal of this section is to establish the existence of a unique
local geodesic flow and local well-posedness for both equations.
Section 3 is about the curvature of the product group associated
with the 2HS. In Section 4 we express the sectional curvature for
2$\mu$HS in terms of its Christoffel map and present a large
subspace of positive sectional curvature.
\section{Geometric aspects of $2$HS and 2$\mu$HS}
Let us denote by $H^s\Diff(\S)$ the group of
orientation-preserving $H^s$ diffeomorphisms of $\S=\R/\Z$ where
$H^s=H^s(\S)$ is the Sobolev space of order $s>0$ on the circle.
Note that, for any $s>3/2$, $H^s\Diff(\S)$ is a topological group
and a smooth Hilbert manifold modelled on the space $H^s$, but not
a Lie group\red{, \cite{L07',M02}}. We have $TH^s\Diff(\S)\simeq H^s\Diff(\S)\times
H^s(\S)$.
\subsection{The HS equation}
\indent Let us rewrite the HS equation as an autonomous system in terms of
the local flow $X(t)=(\phi(t),\phi_t(t))\in TH^s\Diff(\S)$,
$s>5/2$, for the time-dependent $H^s$ vector field $u(t,\cdot)$ on
$\S$. The usual starting point is to compute $\phi_{tt}$. By the
chain rule and from the relation $\phi_t=u\circ\phi$, we obtain
$$\phi_{tt}=(u_t+uu_x)\circ\phi.$$
We now have to replace $u_t$ by using \autoref{HS}. But since
\autoref{HS} only includes $u_{txx}$, we differentiate twice with
respect to $x$ to obtain
\bea\partial_x^2(u_{t}+uu_x)&=&u_{txx}+3u_xu_{xx}+uu_{xxx}\nonumber\\
&=&u_xu_{xx}\nonumber\\
&=&\frac{1}{2}\partial_xu_x^2.\label{comp1}\eea
Now we are in need of the inverse of the operator
$A=-\partial_x^2$.
\lem\label{lem_inverse} Let $A$ be the operator $-\partial_x^2$
with domain
$$D(A)=\set{f\in H^s(\S)}{f(0)=0},\quad s>5/2.$$
For any $s>5/2$, $A$ is a topological isomorphism
$$D(A)\to\set{f\in H^{s-2}(\S)}{\int_\S f(x)\dx=0}$$
with the inverse
\bea(A^{-1}f)(x)=-\int_0^x\int_0^yf(z)\d z\d
y+x\int_\S\int_0^yf(z)\d z\d y.\nonumber\eea
\endlem\rm
By Sobolev's imbedding theorem $u_x\in C(\S)$ and hence the
right-hand side of \eqref{comp1} is a function with zero mean.
Thus it is in the domain of $A^{-1}$ and we conclude
$$\phi_{tt}=-\frac{1}{2}\left[A^{-1}\partial_x(\phi_t\circ\phi^{-1})_x^2\right]\circ\phi.$$
Note also that $u_t+uu_x$ must belong to the domain of $A$ which
suggests that we will need the assumption $u(0)=0$.
Setting
\bea\label{GammanullHS}\Gamma(u,v)=-\frac{1}{2}A^{-1}(u_xv_x)_x,\eea
we obtain a symmetric bilinear operator $H^s(\S)\times H^s(\S)\to
H^s(\S)$. In fact, this \emph{Christoffel map} is smooth which
enables the following geometric approach, established in \cite{L08}:\\
\indent
Let $\text{Rot}(\S)\subset H^s\Diff(\S)$ be the subgroup of
rotations $x\mapsto x+d$ for some $d\in\R$. We denote by
$H^s\Diff(\S)/\text{Rot}(\S)$ the space of right cosets
$\text{Rot}(\S)\circ\phi=\set{\phi(\cdot)+d}{d\in\R}$ for $\phi\in
H^s\Diff(\S)$ and set $M^s=\set{\phi\in H^s\Diff(\S)}{\phi(0)=0}$.
We have
$$M^s=\set{\id+u}{u\in H^s,\,u_x>-1,\,u(0)=0}$$
and thus $M^s$ is an open subset of the closed hyperplane
$$\id+E^s:=\id+\set{u\in H^s}{u(0)=0}\subset H^s.$$
Writing the elements of $H^s\Diff(\S)/\text{Rot}(\S)$ as $[\phi]$,
the map $[\phi]\mapsto \phi-\phi(0)$ establishes a diffeomorphism
$H^s\Diff(\S)/\text{Rot}(\S)\to M^s$, showing in this way that
$M^s$ is a global chart for $H^s\Diff(\S)/\text{Rot}(\S)$.
Furthermore, all tangent spaces $T_\phi M^s$ can be identified
with $E^s$. Next, we define a right-invariant metric on
$H^s\Diff(\S)/\text{Rot}(\S)$ by setting
\bea\label{HSmetric}\ska{U}{V}_\phi=\frac{1}{4}\int_\S(U\circ\phi^{-1})
A(V\circ\phi^{-1})\dx=\frac{1}{4}\int_\S\frac{U_xV_x}{\phi_x}\dx\eea
for tangent vectors $U,V\in T_\phi M^s\simeq E^s$ at $\phi\in
M^s$. Recall that the bilinear form
$\ska{\cdot}{\cdot}_{\text{id}}$ at the identity, induced by the
operator $A$ defined in Lemma~\ref{lem_inverse}, is the $\dot
H^1$-metric and that our definition of $A$ ensures that
$\ska{\cdot}{\cdot}_{\text{id}}$ is indeed a positive definite
inner product\footnote{The factor $1/4$ is introduced to obtain
that the sectional curvature for HS is identically equal to one.}.
Furthermore the metric \eqref{HSmetric} is compatible with the
affine connection $\nabla$ defined locally by
\beq\label{connectionHS}\nabla_XY(\phi)=DY(\phi)\cdot
X(\phi)-\Gamma(\phi;Y(\phi),X(\phi)),\eeq
where
$\Gamma(\phi;\cdot,\cdot)=R_\phi\circ\Gamma(\id,\cdot,\cdot)\circ
R_\phi^{-1}$ is the smooth Christoffel map for the HS equation
with $\Gamma(\id,u,v)=-\frac{1}{2}A^{-1}(u_xv_x)_x$. As proved in
\cite{L08}, the geodesics of the $\dot H^1$ right-invariant metric
are described by the HS equation: Let $J\subset\R$ be an open
interval and let $\phi\colon J\to H^s\Diff(\S)$ be a smooth curve.
Then the curve $u\colon J\to T_{\text{id}}H^s\Diff(\S)$ defined by
$u\colon t\mapsto\phi_t\circ\phi^{-1}$ satisfies the HS equation
\eqref{HS} if and only if the curve $[\phi]\colon J\to
H^s\Diff(\S)/\text{Rot}(\S)$ given by $[\phi]\colon
t\mapsto[\phi(t)]$ is a geodesic with respect to $\nabla$. The
geodesics in $H^s\Diff(\S)/\text{Rot}(\S)$ can be found explicitly
by the method of characteristics: For $u_0\in T_{\text{id}}M^s$
with $\ska{u_0}{u_0}=1$ the unique geodesic
$\phi\colon[0,T^*(u_0))\to M^s$ with $\phi(0)=\id$ and
$\phi_t(0)=u_0$ is given by
$$\phi(t)=\id-\frac{1}{8}\left(A^{-1}\partial_x\left(u_{0x}^2\right)\right)(1-\cos 2t)+\frac{1}{2}u_0\sin 2t,$$
where the maximal time of existence is
$$T^*(u_0)=\frac{\pi}{2}+\arctan\left(\frac{1}{2}\min_{x\in\S}u_{0x}(x)\right)<\pi/2.$$
Observe that the associated solution $u=\phi_t\circ\phi^{-1}\in
C([0,T^*);E^s)\cap C^1([0,T^*);E^{s-1})$ of the HS is not unique;
the set of solutions is
$$\{t\mapsto u(t,\cdot-c(t))+c'(t)\}\subset C([0,T);H^s(\S))\cap C^1([0,T);H^{s-1}(\S)),$$
where $T\leq T^*$ is the maximal time of existence,
$c\colon[0,T)\to\R$ is an arbitrary $C^1$-function with
$c(0)=c'(0)=0$ and if $T<T^*$, then $|c(t)|\to\infty$ as $t\to T$
from below.
Further geometric aspects of the HS equation are discussed in
\cite{L07',L07''}.
\subsection{The $\mu$HS equation}
The inertia operator for $\mu$HS is $A=\mu-\partial_x^2$, where $\mu(u)=\int_\S u\dx$. The following lemma can be found in \cite{LMT09}.

\lem For $s\geq 2$, the linear operator $A=\mu-\partial_x^2:
H^s\to H^{s-2}$ is a topological isomorphism with the inverse
\bea(A^{-1}f)(x)&=&\left(\frac{1}{2}x^2-\frac{1}{2}x+
\frac{13}{12}\right)\int_0^1f(a)\d a+\left(x-\frac{1}{2}\right)
\int_0^1\int_0^af(b)\d b\d a\nonumber\\
&&-\int_0^x\int_0^af(b)\d b\d a+\int_0^1\int_0^a\int_0^bf(c)\d c\d
b\d a\nonumber\eea
and the Green's function
$$(A^{-1}f)(x)=\int_\S g(x-x')f(x')\d x',\quad g(x-x')=\frac{1}{2}(x-x')^2-\frac{1}{2}|x-x'|+\frac{13}{12}.$$
\endlem\rm
Let $s>5/2$. The Christoffel operator $\Gamma=\Gamma_\id$ for
\autoref{muHS} is
\beq\Gamma(u,v)=-A^{-1}\left(\mu(u)v+\mu(v)u+\frac{1}{2}u_xv_x\right)_x,\label{ChristoffelmuCH}\eeq
since $\mu$HS can be written as $u_t+uu_x=\Gamma(u,u)$.
The bilinear map
$$\ska{U}{V}_\phi=\mu(U\circ\phi^{-1})\mu(V\circ\phi^{-1})+\int_\S \frac{U_xV_x}{\phi_x}\dx,\quad
U,V\in T_\phi H^s\Diff(\S)$$
defines a right-invariant inner product on $H^s\Diff(\S)$ and the
pair $(H^s\Diff(\S),\ska{\cdot}{\cdot})$ is a Riemannian manifold.
The $\mu$HS Christoffel map depends smoothly on $\phi$ and defines
a Riemannian covariant derivative on $H^s\Diff(\S)$ via
\eqref{connectionHS}, compatible with the right-invariant metric
$\ska{\cdot}{\cdot}$. In consequence, the $\mu$HS possesses a
unique geodesic flow $\phi\in H^s\Diff(\S)$. As a geodesic
equation on $H^s\Diff(\S)$, the $\mu$HS reads
$\phi_{tt}=\Gamma_\phi(\phi_t,\phi_t)$ in Lagrangian coordinates.
We also conclude that $\mu$HS is locally well-posed in $H^s$ for
$s>5/2$, cf.~\cite{KLM08}.
\subsection{Generalities on semidirect products}
Let $G$ be a Lie group and $V$ be a vector space. If $G$ acts on
the right on $V$, one defines
$$(g_1,v_1)(g_2,v_2)=(g_1g_2,v_2+v_1g_2)$$
and with this product, $G\times V$ becomes a Lie group (the \emph{semidirect product of $G$ and $V$}) which is denoted as $G\circledS V$. It is easy to see
that $(e,0)$ is the neutral element, where $e$ denotes the neutral
element of $G$, and that $(g,v)$ has the inverse
$(g^{-1},-vg^{-1})$. To obtain the Lie bracket on the Lie algebra
$\g\circledS V$, we consider the inner automorphism
$$I_{(g,v)}(h,w)=(g,v)(h,w)(g,v)^{-1}=(ghg^{-1},-vg^{-1}+(w+vh)g^{-1}).$$
Writing $v\xi$ for the induced infinitesimal action of $\g$ on
$V$, i.e., the map
$$V\times\g\mapsto V,\quad(v,\xi)\mapsto v\xi:=\left.\frac{\d}{\d t}vg(t)\right|_{t=0},$$
$g(t)$ being a curve in $G$ starting from $e$ in the direction of
$\xi$, we obtain
$$\Ad_{(g,v)}(\xi,w)=(\Ad_g\xi,(w+v\xi)g^{-1}),$$
$$\ad_{(\eta,v)}(\xi,w)=(\ad_\eta\xi,v\xi-w\eta)$$
and hence
$$[(\xi_1,v_1),(\xi_2,v_2)]=\ad_{(\xi_2,v_2)}(\xi_1,v_1)=([\xi_1,\xi_2],v_2\xi_1-v_1\xi_2).$$
\indent For our purposes, we consider the semidirect product of
the orientation-preserving diffeomorphisms $H^s\Diff(\S)$ with
$H^{s-1}$; the structure of our equations motivates to enforce the
second component to have one order less regularity than the first,
cf. also \cite{EKL10}. We will use the notation $H^sG$ and
$H^sG_0$ for the Lie groups $H^s\Diff(\S)\circledS H^{s-1}(\S)$
and $[H^s\Diff(\S)/\text{Rot}(\S)]\circledS E^{s-1}(\S)$. The
group product in these groups is defined by
$$(\phi_1,f_1)(\phi_2,f_2):=(\phi_1\circ\phi_2,f_2+f_1\phi_2)$$
where $\circ$ denotes the group product in $H^s\Diff(\S)$ (i.e.,
composition) and $f\phi:= f\circ\phi$ is a right action of
$H^s\Diff(\S)$ on the scalar functions on $\S$. The neutral
element is $(\id,0)$ and $(\phi,f)$ has the inverse
$(\phi^{-1},-f\circ\phi^{-1})$. The above calculations show that
$$\Ad_{(\phi,f)}(u,\rho)=(\Ad_\phi u,(f_{x}u+\rho)\circ\phi^{-1}),$$
$$\ad_{(v,f)}(u,\rho)=(\ad_vu,f_xu-\rho_xv)$$
and
$$[(u_1,u_2),(v_1,v_2)]=([u_1,v_1],v_{2x}u_1-u_{2x}v_1),$$
where $[u_1,v_1]=v_{1x}u_1-u_{1x}v_1$ is the Lie bracket induced
by right-invariant vector fields on $H^s\Diff(\S)$. Observe that
$TH^sG\simeq H^sG\times (H^s\times H^{s-1})$ and $TH^sG_0\simeq
H^sG_0\times (E^s\times E^{s-1})$. For further details about
semidirect product groups we refer to \cite{HMR98,HT09}.
\subsection{The 2HS equation} Inspired by the results for 2CH and
2DP in \cite{EKL10}, we work with the configuration space
$$H^sG_0=[H^s\Diff(\S)/\text{Rot}(\S)]\circledS E^{s-1}(\S),\quad s>5/2.$$
The Lagrangian for 2HS motivates to define the right-invariant
metric on $H^sG_0$ which equals the $\dot H^1$-metric for the
first component plus the $L_2$-metric for the second component at
the identity $(\id,0)$:
\beq\label{metric2HS}\ska{U}{V}_{(\phi,f)}=\ska{u_1\circ\phi^{-1}}{v_1\circ\phi^{-1}}_{\dot
H^1}+\ska{u_2\circ\phi^{-1}}{v_2\circ\phi^{-1}}_{L_2},\eeq
for $U=(u_1,u_2),V=(v_1,v_2)\in T_{(\phi,f)}H^sG_0$.
Let $A$ and $\Gamma^0$ as in Lemma~\ref{lem_inverse} and \eqref{GammanullHS}. With the 2HS
Christoffel map $\Gamma$ on $E^{s}\times E^{s-1}$ given by
\bea\label{Christoffel2HS}\Gamma(X,Y)=\left(%
\begin{array}{c}
  \Gamma^0_{\text{id}}(X_1,Y_1)-\frac{1}{2}A^{-1}(X_2Y_2)_x \\
  -\frac{1}{2}(X_{1x}Y_2+Y_{1x}X_2) \\
\end{array}%
\right),\eea
we define the map $\Gamma\colon (M^s\times
E^{s-1})\times(E^s\times E^{s-1})^2\to E^s\times E^{s-1}$,
\beq\Gamma_{(\phi,f)}(X,Y)=\Gamma((\phi,f);X,Y)=\Gamma(X\circ\phi^{-1},
Y\circ\phi^{-1})\circ\phi.\label{Christoffel2HS2}\eeq
Finally, we introduce the affine connection
\beq\label{connection}\nabla_XY(\phi,f)=DY(\phi,f)\cdot X(\phi,f)-\Gamma((\phi,f);Y(\phi,f),X(\phi,f)).\eeq
The proof of the following proposition uses standard arguments and
it is omitted for the reader's convenience.
\prop Let $s > 5/2$. Let $H^sG_0 = [H^{s}\Diff(\S)/\text{\rm
Rot}(\S)]\circledS E^{s-1}(\S)$ and let $\Gamma$ be the
Christoffel map defined in \eqref{Christoffel2HS} and
\eqref{Christoffel2HS2}. Then $\Gamma$ defines a smooth spray on
$H^sG_0$, i.e., the map
$$
([\varphi],f)\mapsto \Gamma_{([\varphi],f)}\colon H^sG_0 \to
\mathcal{L}^2_{\text{\rm sym}}(E^{s}\times E^{s-1};E^{s}\times
E^{s-1})
$$
is smooth. Moreover, the metric $\ska{\cdot}{\cdot}$ defined in
\eqref{metric2HS} is a smooth (weak) Riemannian metric on
$H^sG_0$, i.e., the map
$$
([\varphi],f) \mapsto \ska{\cdot}{\cdot}_{([\varphi],f)}\colon
H^sG_0 \to \mathcal L^2_{\text{\rm
sym}}(T_{([\varphi],f)}H^sG_0;\R)
$$
is a smooth section of the bundle $\mathcal{L}^2_{\text{\rm
sym}}(TH^sG_0;\R)$. Finally, the connection $\nabla$ and the
metric $\ska{\cdot}{\cdot}$ are compatible in the sense that
$$
X\ska{Y}{Z}=\ska{\nabla_XY}{Z} + \ska{Y}{\nabla_XZ}
$$
for all vector fields $X,Y, Z$ on $H^sG_0$.
\endprop\rm
As a consequence, we obtain a unique local geodesic flow $(\phi(t),f(t))\in
H^sG_0$ for 2HS satisfying
\beq\label{geoeqn2HS}(\phi_{tt},f_{tt})=\Gamma_{(\phi,f)}((\phi_t,f_t),(\phi_t,f_t)).\eeq
\thm\label{2HSgeodesicth} Let $s > 5/2$. Then there exists an open
interval $J$ centered at $0$ and an open neighborhood $U$ of
$(0,0) \in E^{s}\times E^{s-1}$ such that for each $(u_0,\rho_0)
\in U$ there exists a unique solution $(\varphi, f) \in
C^\infty(J, H^sG_0)$ of \eqref{geoeqn2HS} satisfying
$(\varphi(0),f(0))=(\id,0)$ and $(\varphi_t(0), f_t(0)) =
(u_0,\rho_0)$. Furthermore, the solution depends smoothly on the
initial data in the sense that the local flow $\Phi\colon J \times
U \to H^sG_0$ defined by $\Phi(t, u_0, \rho_0) = (\varphi(t; u_0,
\rho_0), f(t; u_0, \rho_0)) $ is a smooth map.
\endthm\rm
Writing the Cauchy problem for 2HS in the form
\bea\label{weak2HS} \left\{%
\begin{array}{ccc}
  u_t+uu_x       & = & -\frac{1}{2}A^{-1}\left(u_x^2+\rho^2\right)_x,\\
  \rho_t+u\rho_x & = & -\rho u_x,\\
  (u(0),\rho(0)) & = &(u_0,\rho_0),\\
\end{array}
\right.\eea
Theorem~\ref{2HSgeodesicth} implies that \autoref{weak2HS} in
terms of the Euclidean variables $u=\phi_t\circ\phi^{-1}$ and
$\rho=f_t\circ\phi^{-1}$ is locally well-posed.
\cor\label{cor_lwp_2HS} Suppose $s>5/2$. Then for any
$(u_0,\rho_0) \in E^{s} \times E^{s-1}$ there exists an open
interval $J$ centered at $0$ and a unique solution
$$
 (u, \rho) \in C(J, E^{s} \times E^{s-1}) \cap C^1(J, E^{s-1} \times E^{s-2})
$$
of the Cauchy problem \eqref{weak2HS} which depends continuously
on the initial data $(u_0, \rho_0)$.
\endcor\rm
A similar result is true in the $C^n$-category with $n\geq 2$.
\subsection{The 2$\mu$HS equation} We define a right-invariant
metric on $H^sG$, $s>5/2$, which equals the inner product induced
by $\mu-\partial_x^2$ for the first component plus the $L_2$ inner
product for the second one at the identity, i.e.,
\beq\ska{U}{V}_{(\phi,f)}=\mu(u_1\circ\phi^{-1})\mu(v_1\circ\phi^{-1})+\int_\S\frac{u_{1x}v_{1x}}{\phi_x}\dx+\int_\S
u_2v_2\phi_x\dx\label{skapro2muHS}\eeq
for any $U=(u_1,u_2),V=(v_1,v_2)\in T_{(\phi,f)} H^sG$. With
$\Gamma^0$ as in \eqref{ChristoffelmuCH} we define a
right-invariant Christoffel map for 2$\mu$HS by
\bea\label{Christoffel2muHS}\Gamma_{(\text{id},0)}(X,Y)=\left(%
\begin{array}{c}
  \Gamma^0_{\text{id}}(X_1,Y_1)-\frac{1}{2}(\mu-\partial_x^2)^{-1}(X_2Y_{2})_x \\
  -\frac{1}{2}(X_{1x}Y_2+Y_{1x}X_2)\\
\end{array}%
\right).\eea
and an affine connection via \eqref{connection}. Then the
following is easy to derive.
\prop\label{prop_geometry2muHS} Let $s > 5/2$. Let $H^sG =
H^{s}\Diff(\S)\circledS H^{s-1}(\S)$ and let $\Gamma$ be the
Christoffel map defined in \eqref{Christoffel2muHS}. Then $\Gamma$
defines a smooth spray on $H^sG$, i.e., the map
$$
(\varphi,f)\mapsto \Gamma_{(\varphi,f)}\colon H^sG \to
\mathcal{L}^2_{\text{\rm sym}}(H^{s}(\S)\times
H^{s-1}(\S);H^{s}(\S)\times H^{s-1}(\S))
$$
is smooth. Moreover, the metric $\ska{\cdot}{\cdot}$ defined by
\eqref{skapro2muHS} is a smooth (weak) Riemannian metric on
$H^sG$, i.e., the map
$$
(\varphi,f) \mapsto \ska{\cdot}{\cdot}_{(\varphi,f)}: H^sG \to
\mathcal L^2_{\text{\rm sym}}\left(T_{(\varphi,f)}H^sG;\R\right)
$$
is a smooth section of the bundle $\mathcal{L}^2_{\text{\rm
sym}}\left(TH^sG;\R\right)$. Finally, the connection $\nabla$ is a
Riemannian covariant derivative, compatible with
$\ska{\cdot}{\cdot}$.
\endprop\rm
We thus know that the 2$\mu$HS is a reexpression of the geodesic
flow of the connection $\nabla$ defined in \eqref{connection} on
the product $H^sG$. The geodesic equation reads
\eqref{geoeqn2HS}.
We have the following local well-posedness result.
\thm\label{2muHSgeodesicth} Let $s > 5/2$ and let $\Gamma$ be the
2$\mu$HS Christoffel map. Then there exists an open interval $J$
centered at $0$ and an open neighborhood $U$ of $(0,0) \in
H^{s}(\S)\times H^{s-1}(\S)$ such that for each $(u_0,\rho_0) \in
U$ there exists a unique solution $(\varphi, f) \in C^\infty(J,
H^sG)$ of \eqref{geoeqn2HS} satisfying $(\varphi(0),f(0))=(\id,0)$
and $(\varphi_t(0), f_t(0)) = (u_0,\rho_0)$. Furthermore, the
solution depends smoothly on the initial data in the sense that
the local flow $\Phi\colon J \times U \to H^sG$ defined by
$\Phi(t, u_0, \rho_0) = (\varphi(t; u_0, \rho_0), f(t; u_0,
\rho_0)) $ is a smooth map.
\endthm\rm
We write the Cauchy problem for 2$\mu$HS in the form
\bea\label{weak2muHS} \left\{%
\begin{array}{ccc}
  u_t+uu_x       & = & -(\mu-\partial_x^2)^{-1}\left(\frac{1}{2}u_x^2 +2\mu(u)u + \frac{1}{2} \rho^2\right)_x,\\
  \rho_t+u\rho_x & = & -\rho u_x,\\
  (u(0), \rho(0))& = & (u_0, \rho_0).
\end{array}%
\right.\eea
It follows from Theorem~\ref{2muHSgeodesicth} that 2$\mu$HS is
locally well-posed in $H^{s} \times H^{s-1}$ for $s > 5/2$.
\cor\label{2HSlocalwellposedcor} Suppose $s > 5/2$. Then for any
$(u_0,\rho_0) \in H^{s}(\S) \times H^{s-1}(\S)$ there exists an
open interval $J$ centered at $0$ and a unique solution
$$
 (u, \rho) \in C(J, H^{s}(\S) \times H^{s-1}(\S)) \cap C^1(J, H^{s-1}(\S) \times H^{s-2}(\S))
$$
of the Cauchy problem \eqref{weak2muHS} which depends continuously
on the initial data $(u_0, \rho_0)$.
\endcor\rm
The previous results hold with the obvious changes also in the
$C^n$-category, $n\geq 2$.
\section{The curvature of $H^sG_0$ associated with the 2HS}
Let us denote by
$$R(X,Y)Z=\nabla_X\nabla_YZ-\nabla_Y\nabla_XZ-\nabla_{[X,Y]}Z$$
the curvature tensor of $H^sG_0$ equipped with the right-invariant
metric \eqref{metric2HS}. In the following theorem we compute an
explicit formula for $R$ and show that the sectional curvature
$$S(X,Y)=\frac{\ska{R(X,Y)Y}{X}}{\ska{X}{X}\ska{Y}{Y}-\ska{X}{Y}^2}$$
has the constant positive value $1/4$.
\thm\label{lem_seccurvature2HS} The curvature tensor for the 2HS
equation on $H^sG_0$, $s>5/2$, equipped with the right-invariant
metric \eqref{metric2HS}, for vector fields $X,Y,Z$, is given by
$$4R(X,Y)Z=X\ska{Y}{Z}-Y\ska{X}{Z}.$$
In particular, the sectional curvature for 2HS is constant and
equal to $1/4$.
\endthm
\proof
We have the following local formula for $R$ in terms of the
Christoffel map \eqref{Christoffel2HS2}:
$$R(X,Y)Z=
D_1\Gamma_p(Z,X)Y-D_1\Gamma_p(Z,Y)X+\Gamma_p(\Gamma_p(Z,Y),X)-\Gamma_p(\Gamma_p(Z,X),Y),$$
for any vector fields $X,Y,Z$ on $H^sG$, where $D_1$ denotes
differentiation with respect to $p=(\phi,f)$. By right-invariance of
$\Gamma$, i.e.,
$$\Gamma_p(X,Y)\circ\psi=\Gamma_{p\circ\psi}(X\circ\psi,Y\circ\psi),$$
it holds that $$R(X,Y)Z\circ\phi^{-1}=R(u,v)w$$ if $X=u\circ\phi$, $Y=v\circ\phi$ and $Z=w\circ\phi$.
Therefore, it suffices to consider the curvature at
$(\id,0)$. We write $\Gamma=\Gamma_{(\id,0)}$ and denote the
components of $u$ by $u_1$ and $u_2$ and similarly for $v,w$. A
lengthy but straightforward computation shows that
\bea R(u,v)w&=&
D_1\Gamma(w,u)v-D_1\Gamma(w,v)u+\Gamma(\Gamma(w,v),u)-\Gamma(\Gamma(w,u),v),\nonumber\\
&=&-\Gamma(w_xv_1,u)-\Gamma(u_xv_1,w)+\Gamma(w,u)_xv_1\nonumber\\
&&\quad+\Gamma(w_xu_1,v)+\Gamma(v_xu_1,w)-\Gamma(w,v)_xu_1\nonumber\\
&&\quad+\Gamma(\Gamma(w,v),u)-\Gamma(\Gamma(w,u),v).\nonumber\eea
In the first component, we have the terms
\bea
&&-\Gamma^0(w_{1x}v_1,u_1)+\frac{1}{2}A^{-1}(w_{2x}v_1u_2)_x-\Gamma^0(u_{1x}v_1,w_1)+\frac{1}{2}A^{-1}(u_{2x}v_1w_2)_x\nonumber\\
&&+\Gamma^0(w_1,u_1)_xv_1-\frac{1}{2}[A^{-1}(w_2u_2)_x]_xv_1+\Gamma^0(w_{1x}u_1,v_1)-\frac{1}{2}A^{-1}(w_{2x}u_1v_2)_x\nonumber\\
&&+\Gamma^0(v_{1x}u_1,w_1)-\frac{1}{2}A^{-1}(v_{2x}u_1w_2)_x-\Gamma^0(w_1,v_1)_xu_1+\frac{1}{2}[A^{-1}(w_2v_2)_x]_xu_1\nonumber\\
&&+\Gamma^0\left(\Gamma^0(w_1,v_1)-\frac{1}{2}A^{-1}(w_2v_2)_x,u_1\right)+\frac{1}{4}A^{-1}(w_{1x}v_2u_2+v_{1x}w_2u_2)_x\nonumber\\
&&-\Gamma^0\left(\Gamma^0(w_1,u_1)-\frac{1}{2}A^{-1}(w_2u_2)_x,v_1\right)-\frac{1}{4}A^{-1}(w_{1x}u_2v_2+u_{1x}w_2v_2)_x.
\nonumber\eea
Using that
$$\partial_xA^{-1}\partial_x=\mu-1$$
and the relation
$$\Gamma^0(\Gamma^0(w_1,v_1),u_1)-\Gamma^0(\Gamma^0(w_1,u_1),v_1)=-\frac{1}{4}u_1\mu(w_{1x}v_{1x})+\frac{1}{4}v_1\mu(w_{1x}u_{1x}),$$
cf.~\cite{L07'}, we see that these terms equal
\bea&&\frac{1}{2}A^{-1}\partial_x\left[(w_{1x}v_1)_xu_{1x}+(u_{1x}v_1)_xw_{1x}-(w_{1x}u_1)_xv_{1x}-(v_{1x}u_1)_xw_{1x}\right]\nonumber\\
&&-\frac{1}{2}(\mu-1)(w_{1x}u_{1x})v_1+\frac{1}{2}(\mu-1)(w_{1x}v_{1x})u_1\nonumber\\
&&+\frac{1}{2}A^{-1}\left[(w_{2x}v_1u_2)_x+(u_{2x}v_1w_2)_x-(w_{2x}u_1v_2)_x-(v_{2x}u_1w_2)_x\right]\nonumber\\
&&-\frac{1}{2}(\mu-1)(w_2u_2)v_1+\frac{1}{2}(\mu-1)(w_2v_2)u_1\nonumber\\
&&-\frac{1}{4}u_1\mu(w_{1x}v_{1x})+\frac{1}{4}v_1\mu(w_{1x}u_{1x})-\frac{1}{2}A^{-1}\partial_x
\left[\left(-\frac{1}{2}(\mu-1)(w_2v_2)\right)u_{1x}\right]\nonumber\\
&&+\frac{1}{2}A^{-1}\partial_x\left[\left(-\frac{1}{2}(\mu-1)(w_2u_2)\right)v_{1x}\right]
+\frac{1}{4}A^{-1}\partial_x(v_{1x}w_2u_2-u_{1x}w_2v_2).
\label{firstcomponent}\eea
To see that the terms with $A^{-1}\partial_x$ cancel out, we use
that $u_1(0)=v_1(0)=0$ so that
\bea\frac{1}{2}w_{1x}u_{1x}v_1-\frac{1}{2}u_1w_{1x}v_{1x}&=&-A^{-1}\partial_x^2\left(
\frac{1}{2}w_{1x}u_{1x}v_1-\frac{1}{2}u_1w_{1x}v_{1x}\right)\nonumber\\
&=&\frac{1}{2}A^{-1}\partial_x(w_{1xx}(u_1v_{1x}-u_{1x}v_1)\nonumber\\
&&\hspace{2.5cm}+w_{1x}u_1v_{1xx}-w_{1x}u_{1xx}v_{1}),\nonumber\eea
which coincides up to sign with the first row terms in
\eqref{firstcomponent}, and
\bea\frac{1}{2}w_2u_2v_1-\frac{1}{2}w_2v_2u_1&=&-A^{-1}\partial_x^2\left(\frac{1}{2}w_2u_2v_1-\frac{1}{2}w_2v_2u_1\right)\nonumber\\
&=&\frac{1}{2}A^{-1}\partial_x(w_{2x}(v_2u_1-u_2v_1)+w_2(v_{2x}u_1-u_{2x}v_1)\nonumber\\
&&\quad+w_2(v_2u_{1x}-u_2v_{1x})).\nonumber\eea
Using $A^{-1}\partial_x^2v_1=-v_1$ and
$A^{-1}\partial_x^2u_1=-u_1$, the first component terms
\eqref{firstcomponent} thus reduce to
$$\frac{1}{4}u_1(\mu(w_{1x}v_{1x})+\mu(w_2v_2))-\frac{1}{4}v_1(\mu(w_{1x}u_{1x})+\mu(w_2u_2)),$$
which is the desired expression. The second component terms are
\bea&&\frac{1}{2}[(w_{1x}v_1)_xu_2+u_{1x}w_{2x}v_1]+\frac{1}{2}[(u_{1x}v_1)_xw_2+w_{1x}u_{2x}v_1]-\frac{1}{2}v_1[w_{1x}u_2+u_{1x}w_2]_x\nonumber\\
&&-\frac{1}{2}[(w_{1x}u_1)_xv_2+v_{1x}w_{2x}u_1]-\frac{1}{2}[(v_{1x}u_1)_xw_2+w_{1x}v_{2x}u_1]+\frac{1}{2}u_1[w_{1x}v_2+v_{1x}w_2]_x\nonumber\\
&&-\frac{1}{2}(\Gamma_1(w,v)_xu_2+u_{1x}\Gamma_2(w,v))+\frac{1}{2}(\Gamma_1(w,u)_xv_2+v_{1x}\Gamma_2(w,u))\nonumber\\
\label{secondcomponent}\eea
and with $\partial_xA^{-1}\partial_x=\mu-1$ we can simplify the
last row terms
$$\Gamma_1(w,v)_x=\frac{1}{2}w_{1x}v_{1x}-\frac{1}{2}\mu(w_{1x}v_{1x})+\frac{1}{2}w_2v_2-\frac{1}{2}\mu(w_2v_2)$$
and
$$\Gamma_1(w,u)_x=\frac{1}{2}w_{1x}u_{1x}-\frac{1}{2}\mu(w_{1x}u_{1x})+\frac{1}{2}w_2u_2-\frac{1}{2}\mu(w_2u_2).$$
It is now easy to see that the terms in \eqref{secondcomponent}
reduce to
$$\frac{1}{4}u_2(\mu(w_{1x}v_{1x})+\mu(w_2v_2))-\frac{1}{4}v_2(\mu(w_{1x}u_{1x})+\mu(w_2u_2))$$
so that we obtain
$$R(u,v)w=\frac{1}{4}u\ska{v}{w}-\frac{1}{4}v\ska{u}{w}.$$
By the definition of the sectional curvature, we have
$$S(u,v)=\frac{\ska{R(u,v)v}{u}}{\ska{u}{u}\ska{v}{v}-\ska{u}{v}^2}=\frac{1}{4}.$$
\qed
\rem Since Lenells \cite{L07'} uses a different scaling for the
$\dot H^1$-metric, he comes to the result that the sectional
curvature for the HS is identically equal to $1$. Note carefully
that we have only used that $u_1$ and $v_1$ vanish at zero; a
corresponding assumption on the second components is not necessary
in the above proof.
\endrem\rm
\red{The above result is of particular interest concerning the stability of the
geodesic flow associated with the 2HS system, cf. the appendix.
The very recent paper \cite{W11} explains that the geodesic flow for the 2-component
Hunter-Saxton system allows a continuation (beyond the breaking time of the associated
solution $(u,\rho)$ of the original equation) on the space
$\mathcal M^0_{AC}=M^{AC}\circledS H^0(\S)$, where $M^{AC}$ is the set of nondecreasing
absolutely continuous functions $\phi\colon[0,1]\to[0,1]$ with $\phi(0)=0$ and $\phi(1)=1$.
Precisely, the flow variables $(\phi,f)\in\mathcal M^0_{AC}$ for 2HS satisfy the geodesic equation
$(\phi_{tt},f_{tt})=\Gamma_{(\phi,f)}((\phi_t,f_t),(\phi_t,f_t))$ for \emph{any} positive $t$.
Interestingly, the space $M^{AC}$ is bijective to the open subset
$$\mathcal U^0=\set{f\in H^0(\S)}{\Vert f\Vert_{H^0}=1,\,f>0\text{ a.e. on }\S}$$
of the $L_2$ unit sphere via the mapping
$$f\mapsto\left(x\mapsto\phi(x)=\int_0^xf^2(y)\d y\right).$$
Lenells \cite{L07',L07''} used the constance and positivity of the sectional curvature associated with
HS as a motivation to continue the geodesic flow on the infinite-dimensional sphere $M^{AC}$,
keeping in mind the well-known fact that any finite dimensional Riemannian manifold with
constant positive curvature is locally isometric to a sphere. In the interim, Wunsch \cite{W11} already adopted this approach
for the 2HS system so that our work may serve retroactively as a motivation for considering the problem
of extending the flow variables on a suitable configuration space (which in fact works by multiplying
the sphere $M^{AC}$ with $H^0(\S)$).
}
\section{The curvature of $H^sG$ associated with the 2$\mu$HS}
In this section, $R$ and $S$ stand for the curvature tensor and
the unnormalized sectional curvature for the 2$\mu$HS equation.
\thm\label{lem_seccurv2muHS} The unnormalized sectional curvature
$S(u,v)$ for the 2$\mu$HS equation is given by
$$S(u,v)=\ska{\Gamma(u,v)}{\Gamma(u,v)}-\ska{\Gamma(u,u)}{\Gamma(v,v)}-3\mu(u_{1x}v_1)^2.$$
\endthm\rm
\proof The proof is similar to the proof of Proposition 5.1 in
\cite{EKL10}; the single difference is that the additional terms
involving $\Gamma^0$ in (5.4) of \cite{EKL10} do not cancel out
but give $-3\mu(u_{1x}v_1)^2$ as explained in
\cite{KLM08}.\endproof
In the following we write $S_1$, $S_2$ to distinguish between the
sectional curvature for the one-component $\mu$HS and its
two-component extension. In \cite{KLM08}, the authors prove that
$S_1(u_1,v_1)$ is always positive for any two orthonormal vectors
$u_1$ and $u_2$ (with respect to the scalar product induced by
$\mu-\partial_x^2$). Since we have
$$S_2\left(\left(%
\begin{array}{c}
  u_1 \\
  0 \\
\end{array}%
\right),\left(%
\begin{array}{c}
  v_1 \\
  0\\
\end{array}%
\right)\right)=S_1(u_1,v_1),$$
we see that $S_2$ is positive in the $H^s\Diff(\S)$-direction. To
find a large subspace of positive sectional curvature for $2\mu$HS
with non-trival second component we compute $S_2(u,v)$ for
$$
u=\left(%
\begin{array}{c}
  \cos k_1x \\
  \cos k_2x \\
\end{array}%
\right),\quad
v=\left(%
\begin{array}{c}
  \cos l_1x \\
  \cos l_2x \\
\end{array}%
\right),
$$
where $k_i\neq l_i\in 2\pi\N$, $i=1,2$; this is inspired by
calculations in \cite{EKL10} and \cite{LMP09} where the authors
show that the sectional curvature for CH is positive for any pair
of trigonometric functions. Note that
\bea S_2(u,v)
&=&S_1(u_1,v_1)+\frac{1}{4}\int_\S(u_2v_2)_x(\mu-\partial_x^2)^{-1}(u_2v_2)_x\dx-\int_\S\Gamma^0(u_1,v_1)(u_2v_2)_x\dx\nonumber\\
&&+\frac{1}{4}\int_\S(u_{1x}v_2+v_{1x}u_2)^2\dx-\frac{1}{4}\int_\S(u_2^2)_x(\mu-\partial_x^2)^{-1}(v_2^2)_x\dx\nonumber\\
&&+\frac{1}{2}\int_\S\Gamma^0(u_1,u_1)(v_2^2)_x\dx+\frac{1}{2}\int_\S\Gamma^0(v_1,v_1)(u_2^2)_x\dx-\int_\S
u_{1x}u_2v_{1x}v_2\dx\nonumber\\
&=&S_1(u_1,v_1)+\sum_{j=1}^4I_j,\nonumber\\
\label{S2}\eea
where
\bea
I_1&=&\frac{1}{4}\int_\S(u_2v_2)_x(\mu-\partial_x^2)^{-1}(u_2v_2)_x\dx,\nonumber\\
I_2&=&-\frac{1}{4}\int_\S(u_2^2)_x(\mu-\partial_x^2)^{-1}(v_2^2)_x\dx,\nonumber\\
I_3&=&-\int_\S\Gamma^0(u_1,v_1)(u_2v_2)_x\dx+\frac{1}{2}\int_\S\Gamma^0(u_1,u_1)(v_2^2)_x\dx\nonumber\\
&&\qquad
+\frac{1}{2}\int_\S\Gamma^0(v_1,v_1)(u_2^2)_x\dx,\nonumber\\
I_4&=&\frac{1}{4}\int_\S(u_{1x}v_2+v_{1x}u_2)^2\dx-\int_\S
u_{1x}u_2v_{1x}v_2\dx.\nonumber\eea
We write $A=\mu-\partial_x^2$ and apply the identity
$$A^{-1}\partial_x^2=\partial_x
A^{-1}\partial_x=\partial_x^2A^{-1}=\mu-1.$$
Using integration by parts and the orthogonality relations for
trigonometric functions we find
\bea S_1(u_1,v_1)&=&\ska{\Gamma^0(u_1,v_1)}{\Gamma^0(u_1,v_1)}
-\ska{\Gamma^0(u_1,u_1)}{\Gamma^0(v_1,v_1)}-3\mu(u_{1x}v_1)^2\nonumber\\
&=&-\frac{1}{2}\int_\S
A^{-1}[\partial_x(u_{1x}v_{1x})]A\Gamma^0(u_1,v_1)\dx\nonumber\\
&&\qquad +\frac{1}{2}\int_\S
A^{-1}[\partial_x(u_{1x}^2)]A\Gamma^0(v_1,v_1)\dx\nonumber\\
&=&\frac{1}{2}\int_\S u_{1x}v_{1x}\Gamma^0(u_1,v_1)_x\dx-\frac{1}{2}\int_\S u_{1x}^2\Gamma^0(v_1,v_1)_x\dx\nonumber\\
&=&-\frac{1}{4}\int_\S u_{1x}v_{1x}(A^{-1}\partial_x^2)(u_{1x}v_{1x})\dx+\frac{1}{4}\int_\S u_{1x}^2(A^{-1}\partial_x^2)(v_{1x}^2)\dx\nonumber\\
&=&\frac{1}{4}\mu(u_{1x}^2)\mu(v_{1x}^2)\nonumber\\
&=&\frac{1}{16}k_1^2l_1^2.\nonumber\\\label{S1}\eea
Our choice of $k_1$ and $l_1$ implies that the one-component
sectional curvature is strictly positive. All we have to show is
that the second component terms do not contribute negative terms
which make the total sectional curvature negative. Similar
computations show that the terms $I_1$ and $I_2$ in \eqref{S2} are
$$I_1=-\frac{1}{4}\int_\S u_2v_2\partial_x^2A^{-1}u_2v_2\dx=\frac{1}{4}\int_\S u_2v_2(1-\mu)(u_2v_2)\dx=\frac{1}{4}\int_\S
u_2^2v_2^2\dx$$
and
$$I_2=\frac{1}{4}\int_\S u_2^2\partial_x^2A^{-1}v_2^2\dx=\frac{1}{4}\int_\S u_2^2(\mu-1)v_2^2\dx=-\frac{1}{4}\int_\S u_2^2v_2^2\dx+\frac{1}{16}.$$
Since
\bea-\int_\S\Gamma^0(u_1,v_1)(u_2v_2)_x\dx&=&\frac{1}{2}\int_\S
A^{-1}(u_{1x}v_{1x})_x(u_2v_2)_x\dx\nonumber\\
&=&\frac{1}{2}\int_\S[(1-\mu)(u_{1x}v_{1x})]u_2v_2\dx\nonumber\\
&=&\frac{1}{2}\int_\S u_{1x}u_2v_{1x}v_2\dx\nonumber\eea
we find that
\bea
I_3+I_4&=&\frac{1}{2}\int_\S\Gamma^0(u_1,u_1)(v_2^2)_x\dx+\frac{1}{2}\int_\S\Gamma^0(v_1,v_1)(u_2^2)_x\dx\nonumber\\
&&\quad+\frac{1}{4}\int_\S(u_{1x}^2v_2^2+
v_{1x}^2u_2^2)\dx\nonumber\\
&=&\frac{1}{4}\mu(u_{1x}^2)\mu(v_2^2)+\frac{1}{4}\mu(v_{1x}^2)\mu(u_2^2)
\nonumber\\
&=&\frac{1}{16}(k_1^2+l_1^2).\nonumber\eea
It follows from \eqref{S2} and \eqref{S1} that
$$S_2(u,v)=\frac{1}{16}\left(1+k_1^2+l_1^2+k_1^2l_1^2\right)>\frac{1}{16}.$$
Our calculation also shows that the sectional curvature is equal
to $1/16$ in the direction of the second component since
$$S_2\left(\left(%
\begin{array}{c}
  0 \\
  u_2 \\
\end{array}%
\right),\left(%
\begin{array}{c}
  0 \\
  v_2 \\
\end{array}%
\right)\right)=I_1+I_2=\frac{1}{16}.$$
We have thus shown the following proposition.
\prop Let $s > 5/2$. Let $S(u,v):=\ska{R(u,v)v}{u}$ be the
unnormalized sectional curvature on $H^sG$ associated with the
2$\mu$HS equation. Then
$$S(u,v) > \frac{1}{16}$$
for all vectors $u, v \in T_{(\id, 0)}H^sG$, of the form
$$
u=\left(%
\begin{array}{c}
  \cos k_1x \\
  \cos k_2x \\
\end{array}%
\right),\quad v=\left(%
\begin{array}{c}
  \cos l_1x \\
  \cos l_2x \\
\end{array}%
\right),\qquad k_i\neq l_i \in \{2\pi, 4\pi, \dots\}.$$
Moreover, the normalized sectional curvature satisfies
$$
\frac{S(u,v)}{\ska{u}{u}\ska{v}{v}-\ska{u}{v}^2} = \frac{1}{4}
$$
for all vectors $u, v \in T_{(\id, 0)}H^sG$ of the form
$$
u=\left(%
\begin{array}{c}
  0 \\
  \cos k_2x \\
\end{array}%
\right),\quad v=\left(%
\begin{array}{c}
  0 \\
  \cos l_2x \\
\end{array}%
\right), \qquad k_2\neq l_2 \in \{2\pi, 4\pi, \dots\}.
$$
\endprop\rm
\red{It is explained in \cite{KLM08} that the first-component configuration space $H^s\Diff(\S)$ can be thought of as a solid torus with a cross-section isomorphic to $H^s\Diff(\S)/\S$. The fact that the one-component sectional curvature is positive is proved by a decomposition of the tangent space $T_{\id}H^s\Diff(\S)=U\oplus V$ corresponding to the decomposition $u=\tilde u+\mu(u)$, where $\mu(\tilde u)=0$, i.e., $U$ are the zero mean functions and $V\simeq\R$ are the constants. It is an open problem and a task for further research which geometric interpretations for the group $H^sG$ associated with the $2\mu$HS system can be given and which conclusions for the (non)existence of solutions of 2$\mu$HS can be drawn. While the present section shows the existence of subspaces of positive sectional curvature for 2$\mu$HS, one could ask whether $S_2$ is always positive, for arbitrary second component functions, or whether there are directions of strictly negative sectional curvature.
}
\red{
\section{Appendix: Generalities on geodesic flows on infinite dimensional Lie groups and their stability properties}\label{sec_appendix}
In this appendix, we will survey the most important results of the seminal papers \cite{A66,EM70} which are relevant for the purposes of the paper at hand. It goes back to Arnold's work \cite{A66} to model both Euler's equation for a rotating rigid body and Euler's equation for an ideal fluid on a Lie group with an invariant metric. The Lie group for the rotating rigid body is the matrix group $SO(3)$, whereas the motion of an ideal fluid is modeled on the diffeomorphism group $\Diff(M)$ of volume preserving diffeomorphisms of a certain manifold $M$. While the matrix group $SO(3)$ has finite dimension and is equipped with a \emph{left}-invariant metric, the group $\Diff(M)$ is an infinite dimensional Lie group which is equipped with a \emph{right}-invariant metric. The geometric viewpoint is not only aesthetically appealing, but is also very useful for the study of well-posedness and stability issues. In view of the results of the present paper, we provide a general overview about the geometric picture for ideal fluids (which corresponds to a right-invariant formulation), with a focus on the stability of the geodesic flow.

Let $G$ be a (not necessarily finite dimensional) Lie group with Lie algebra $\g\simeq T_eG$, where $e$ denotes the unit element. We assume that there is an invertible linear operator $A\colon\g\to\g^*$ which is, for historical reasons, going back to Euler's work on the rigid body motion, called an \emph{inertia operator}. We also assume that $\g^*\simeq\g$ (which can often be achieved by considering a suitable subspace of $\g^*$) so that $A$ can in fact be regarded as an automorphism of $\g$. Let $R_g\colon G\to G$ denote the right translation map on $G$. We obtain a right-invariant metric $\rho_A$ on $G$ by setting
$$\rho_A(u,v)=(A[DR_{g^{-1}}u],DR_{g^{-1}}v),$$
for all $u,v\in T_gG$, where $(\cdot,\cdot)$ denotes the dual pairing on $\g^*\times\g$. If $G$ is infinite dimensional, the map $\rho_A$ defines in general only a \emph{weak} Riemannian metric on $G$, i.e., the natural topology on any tangent space $T_gG$ is stronger than the topology induced by the metric $\rho_A$, cf.~\cite{EM70}. Let $\ad_u^*$ denote the dual operator (with respect to $\rho_A$) of the natural action of the Lie algebra on itself given by $\ad_u\colon\g\to\g$, $v\mapsto[u,v]$, and define the bilinear and symmetric map
$$B_e\colon\g\times\g\to\g,\quad B_e(u,v)=\frac{1}{2}\left(\ad_u^*v+\ad_v^*u\right).$$
Next, we introduce an affine connection on $G$ given by
\beq\nabla_{\xi_u}\xi_v=\frac{1}{2}[\xi_u,\xi_v]+B(\xi_u,\xi_v),\label{defcon}\eeq
where $\xi_u$ is the right-invariant vector field on $G$ with value $u$ at $e$ and $B$ denotes the right-invariant tensor field with value $B_e$ at the identity. Let $g(t)$ be a smooth part in $G$ and define its \emph{Eulerian velocity}, which lies in the Lie algebra $\g$, by
$$u(t)=DR_{g^{-1}(t)}\dot g(t).$$
The crucial point is that $g(t)$ is a geodesic for the connection $\nabla$ if and only if its Eulerian velocity satisfies the Euler equation
$$u_t=-B(u,u);$$
see \cite{EK09} for instance.

Interestingly the above formalism also works the other way round: Starting from an Euler equation $u_t=-B(u,u)$ with quadratic right-hand side, defined on the Lie algebra $\g$ of some Lie group $G$, and defining the connection $\nabla$ in terms of the operator $B$ as in \eqref{defcon}, one sees that the Euler equation re-expresses a geodesic flow on the Lie group $G$. Nevertheless, it is \emph{not} clear that there is a right-invariant metric $\rho_A$ on $G$, induced by some inertia operator $A$, such that the connection $\nabla$ is compatible with the metric in the sense that
$$X\left(\rho_A(Y,Z)\right)=\rho_A(\nabla_XY,Z)+\rho_A(\nabla_XZ,Y),$$
for vector fields $X,Y,Z$ on $G$. Recall that the equations under discussion in the main body of this paper are \emph{metric} in the sense that they allow for a Riemannian structure.

Let $x(t)$ denote a geodesic in $G$ and consider the geodesic variation $x(t,s)$ with the associated variation vector field
$$\left.\frac{\text{d} x(t,s)}{\text{d} s}\right|_{s=0}=\xi(t)\in T_{x(t)}G.$$
It is well-known that $\xi$ is a solution of the Jacobi equation
$$\frac{D^2\xi}{Dt^2}=-R(\xi,v)v,$$
where $D/Dt$ is the covariant derivative, $R$ is the curvature tensor associated with the connection $\nabla$ and $v=\dot x(t)$ is the velocity field. By a decomposition of the variation vector $\xi$ into components parallel and perpendicular to the velocity $v$, Arnold showed that the Jacobi equation for the perpendicular component (which is henceforth also denoted as $\xi$ for simplicity) can be written in the form
$$\frac{D^2\xi}{Dt^2}=-\text{grad } U,\quad U=\frac{S}{2}\rho_A(\xi,\xi)\rho_A(v,v);$$
here $S$ is the sectional curvature of the two-dimensional subspace of $T_{x(t)}G$ spanned by $v$ and $\xi$. If the norm of $v$ is equal to $1$ (which can be achieved by a parametrization of the geodesic by arc length), the Jacobi equation for $\xi$ reduces to the harmonic oscillator equation with the potential energy $U$ equal to the product of the curvature in the direction spanned by the velocity vector and the normal component of the variation with the square of length of this normal component. From this, it is obvious that $S<0$ implies an exponential divergence of the geodesics starting near $x(0)$, whereas for $S>0$, convergence of the nearby geodesis is expected; cf.~\cite{A89} for further details. This motivates our research for subspaces of positive sectional curvature in the main body of the paper. Note that curvature computations for geometric evolution equations have a long tradition: They have already been carried out in the 1980's \cite{F88,McK82} and in Misio{\l}ek's paper \cite{M98} about the Camassa-Holm equation; see also \cite{L07} for a more recent paper.
}
\\[.25cm]
\emph{Acknowledgements.} The author thanks Jonatan Lenells (Baylor University, Waco) and Joachim Escher (Leibniz University, Hannover) for bringing the above problems to his attention. \red{A cordial thank for useful remarks that helped to improve the manuscript goes to the anonymous referees}.

\end{document}